\documentclass[sigconf,screen,nonacm=true]{acmart}

\AtBeginDocument{%
  }

\usepackage{fontawesome}

\usepackage{xspace}
\xspaceaddexceptions{\%}
\xspaceremoveexception{-}

\usepackage{colortbl}
\usepackage{xcolor}
\definecolor{gray}{rgb}{0.5,0.5,0.5}
\definecolor{mauve}{rgb}{0.57, 0.37, 0.43}
\definecolor{rltred}{rgb}{0.5,0,0}
\definecolor{rltgreen}{rgb}{0,0.5,0}
\definecolor{rltblue}{rgb}{0,0,0.5}
\definecolor{webgreen}{rgb}{0, 0.5, 0} 
\definecolor{webblue}{rgb}{0, 0, 0.5} 
\definecolor{webred}{rgb}{0.5, 0, 0} 
\definecolor{light-gray}{HTML}{B3B3B3}
\definecolor{dim-gray}{HTML}{696969}
\definecolor{white-close}{HTML}{E2E2E2}
\definecolor{JIAWEIdraft}{rgb}{0.24, 0.6, 0.45}
\definecolor{light-red}{rgb}{0.968627451,0.8078431373,0.8039215686} 
\definecolor{light-green}{rgb}{0.8392156863,0.9921568627,0.8156862745} 
\definecolor{light-gray}{rgb}{0.8745098039,0.8745098039,0.8745098039} 
\definecolor{defcolor}{HTML}{2B7A78} 

\usepackage{graphicx}
\usepackage{subcaption}
\usepackage{booktabs}
\usepackage{multirow}
\usepackage{tabularx}
\usepackage[hang,flushmargin]{footmisc} 

\usepackage{amsfonts}
\usepackage{amsmath}
\usepackage{amsthm}

\usepackage{enumitem}
\setlist[enumerate]{leftmargin=*}
\setlist[itemize]{leftmargin=*}

\usepackage{url}
\usepackage[capitalise,noabbrev]{cleveref} 
\crefname{lstlisting}{listing}{listings}
\Crefname{lstlisting}{Listing}{Listings}

\usepackage{listings}
\lstdefinestyle{qasm}{
  language={},
  morekeywords={OPENQASM, include, def, qubit, bit, measure, h, cx, x, ry, rz, if, return, gate, input, output, int, float},
  sensitive=true,
  morecomment=[l]{//},
  morestring=[b]",
  literate={~}{{\textasciitilde}}1,
  aboveskip=0mm,
  belowskip=0mm,
  columns=flexible,
  basicstyle={\scriptsize\ttfamily},
  numbers=left,
  numberstyle=\tiny\color{gray},
  keywordstyle=\color{blue},
  commentstyle=\color{webgreen},
  stringstyle=\color{mauve},
  breaklines=true,
  breakatwhitespace=true,
  tabsize=2,
  xleftmargin=2em,
  xrightmargin=0pt,
  breakindent=0pt,
  resetmargins=true,
  breakautoindent=false
}

\lstdefinestyle{cli}{
  language={},
  keywordstyle={},
  commentstyle={},
  stringstyle={},
  basicstyle={\scriptsize\ttfamily},
  frame=tb,
  numbers=none,
  aboveskip=1mm,
  belowskip=1mm,
  xleftmargin=0em,
  literate=
    {✓}{{$\checkmark$}}1
    {≈}{{$\approx$}}1
}

\setcopyright{acmlicensed}
\copyrightyear{2018}
\acmYear{2018}
\acmDOI{XXXXXXX.XXXXXXX}
\acmConference[Conference acronym 'XX]{Make sure to enter the correct
  conference title from your rights confirmation email}{June 03--05,
  2018}{Woodstock, NY}
\acmISBN{978-1-4503-XXXX-X/2018/06}

\begin{document}

\title{QUTest: A Native Testing Framework for Quantum Programs}

\author{José Campos}
\email{jcmc@fe.up.pt}
\orcid{0000-0001-7565-8382}
\affiliation{%
  \institution{Faculdade de Engenharia da Universidade do Porto, Porto, Portugal}
  \city{}
  \country{}
}
\affiliation{%
  \institution{LASIGE, Faculdade de Ciências, Universidade de Lisboa, Lisboa, Portugal}
  \city{}
  \country{}
}

\renewcommand{\shortauthors}{Campos}

\begin{abstract}
Quantum programs are often shared as OpenQASM~3 circuits, but tests are still written in host languages such as Python with Qiskit.
We present \textsc{QUTest}, a native framework in which both programs and tests are standard \texttt{.qasm} files.
Tests follow the \emph{Arrange\,/\,Act\,/\,Assert} pattern, while configuration, runtime requirements, and assertions are encoded as pragma comments (\texttt{//\%}), preserving compatibility with existing OpenQASM tools.
\textsc{QUTest} provides 12~assertion types spanning deterministic, statistical, quantum-state, and structural checks, plus a linter and an environment-aware mode for running the same test across selected runtime versions in isolated environments.
Its CLI supports automatic test discovery, runtime compatibility checks, and XML reports for continuous integration.
We describe the pragma language, implementation, and a planned evaluation using coverage and mutation testing.

\noindent
\textsc{QUTest} is available at \url{https://github.com/QBugs/qutest}.

\noindent
Video demo: \url{https://youtu.be/FvgvsiAXuW0}.
\end{abstract}

%
%

\maketitle


\section{Introduction}
\label{sec:introduction}

Software testing is a cornerstone of classical software engineering.
Testing frameworks such as JUnit for Java and
PyTest for Python,
allow developers to write and execute tests in the same language as their production code.
This co-location principle---that code and tests share a language, toolchain, and repository---is widely regarded as essential for maintainability and readability~\cite{beck2002tdd}.

Quantum programs are often written in circuit languages, most notably OpenQASM~3~\cite{OpenQASM3}.
OpenQASM~3 provides a framework-neutral textual format for describing quantum circuits, yet such programs still lack a native testing framework.

Today, testing a quantum program written in OpenQASM typically requires a developer to write a separate test harness in, for example, a Python-based SDK such as Qiskit~\cite{aleksandrowicz2019qiskit} and to use the PyTest framework.
The Python test harness in \Cref{lst:intro-bell-pytest} for the Bell-state program in \Cref{lst:intro-bell-qasm} must
(1)~load the program with Qiskit~\cite{aleksandrowicz2019qiskit},
(2)~configure the simulator backend,
(3)~transpile the program,
(4)~execute the program on the simulator,
(5)~extract the measurement counts, and
(6)~apply a statistical test, e.g., a chi-squared goodness-of-fit test at the 95\% confidence level, to those counts.
The result is a test suite in which the \emph{code under test} resides in one language (OpenQASM), whereas the \emph{test harness} resides in another (Python), is coupled to a specific SDK version (Qiskit), and is fragile with respect to API changes that have nothing to do with the quantum algorithm itself (e.g., \cite{cardinal2025qaoaMigration,qiskitIssue12124}).

\begin{lstlisting}[style=qasm,caption={Bell-state program written in OpenQASM~3.},label={lst:intro-bell-qasm},float=t]
OPENQASM 3;
include "stdgates.inc";
qubit[2] q;
bit[2] m;
h q[0];
cx q[0], q[1];
m = measure q;
\end{lstlisting}
\begin{lstlisting}[language={Python},aboveskip=-6mm,belowskip=-5mm,caption={Test harness for \Cref{lst:intro-bell-qasm} written in Python.% and using Python libraries Qiskit SDK, SciPy, and PyTest.
},label={lst:intro-bell-pytest},float=t]
from pathlib import Path
from scipy.stats import chisquare
from qiskit import qasm3, transpile
from qiskit_aer import AerSimulator

def test_distribution():
    # Arrange
    circuit = qasm3.load(Path("bell.qasm"))

    # Act
    backend = AerSimulator(seed_simulator=42)
    transpiled = transpile(circuit, backend=backend)
    result = backend.run(transpiled, shots=10_000).result()
    counts = result.get_counts()

    # Assert
    EXPECTED = {"00": 0.5, "11": 0.5}
    unexpected = sum(v for k, v in counts.items() if k not in EXPECTED)
    assert unexpected == 0
    observed = [counts.get(k, 0) for k in EXPECTED]
    expected = [10_000 * p for p in EXPECTED.values()]
    _, p_value = chisquare(observed, expected)
    assert p_value >= 0.05
\end{lstlisting}

This language gap introduces several problems.
First, readability suffers: understanding a test requires familiarity with both OpenQASM semantics and the Qiskit API.
Second, portability is reduced: tests written against Qiskit do not transfer to Cirq~\cite{cirq}, Pennylane~\cite{pennylane}, or other SDKs without rewriting the host-language glue.
Third, maintenance costs increase: SDK version upgrades can break the test harness even when the quantum algorithm itself is unchanged~\cite{paltenghi2022bugs}.
Fourth, the barrier to entry rises: researchers who write OpenQASM should not need Python expertise merely to verify that a circuit produces the correct output distribution.

We present \textsc{QUTest}, a testing framework that eliminates this gap.
In \textsc{QUTest}, both the software under test and the test cases live in standard \texttt{.qasm} files.
Test are defined as OpenQASM~3 subroutines (\texttt{def}) and follow the familiar \emph{Arrange}/\emph{Act}/\emph{Assert} pattern~\cite{beck2002tdd}: \emph{Arrange} prepares qubits and classical registers, \emph{Act} applies the quantum algorithm and measures, and \emph{Assert} checks the measurement outcomes.
Backend configuration (shots, seed, simulator type, and runtime SDK/version requirements) and assertions (expected output, distribution metrics, entanglement witnesses, and structural budgets) are expressed as pragma comments prefixed with \texttt{//\%}.
Because these pragmas are syntactically valid OpenQASM comments, every \textsc{QUTest} file remains valid OpenQASM~3 syntax.

\smallskip
\noindent
The contributions of this paper are as follows:
\begin{enumerate}[leftmargin=*]
  \item[\small{$\bigstar$}] A pragma-based annotation language for expressing QASM-native quantum tests (\Cref{sec:pragma}).

  \item[\small{$\bigstar$}] The \textsc{QUTest} tool, which implements this language through a command-line workflow for test discovery, static linting, execution, assertion evaluation, environment-aware runtime selection, and CI-oriented reporting (\Cref{sec:architecture}).
\end{enumerate}

\section{Pragma Language Design}
\label{sec:pragma}

The \textsc{QUTest} pragma language extends OpenQASM~3 files with
structured metadata embedded in comments.
Each pragma line begins with the prefix~\texttt{//\%}, followed by a directive.
Pragmas fall into two groups: \emph{configuration pragmas}, which control
how a circuit is executed (\Cref{sec:pragma:conf}), and \emph{assertion
pragmas}, which define pass/fail criteria (\Cref{sec:pragma:assert}).
Annotation-based test metadata is common in classical frameworks.
JUnit uses Java annotations (\texttt{@Test}, \texttt{@BeforeEach}), whereas
PyTest uses decorators (\texttt{@pytest.mark}).

\Cref{lst:bell} shows the Bell-state program, previously presented
in \Cref{lst:intro-bell-qasm}, wrapped in a function named \texttt{bell}
(lines~4--7).
It also shows the same test (\texttt{test\_distribution} in
\Cref{lst:intro-bell-pytest}), written directly in OpenQASM~3 using
\textsc{QUTest}'s pragmas (lines~9--26).
The test first declares the quantum and classical registers (lines~16 and 17),
calls the program under test (lines~20 and 21), and then applies
the same chi-squared goodness-of-fit test at the 95\%
confidence level as the Python baseline (lines~24 and 25).

\begin{lstlisting}[style=qasm,aboveskip=0mm,belowskip=-5mm,caption={Bell-state program and one test case, both written in OpenQASM~3 and using \textsc{QUTest}'s pragmas.},label={lst:bell},float=t]
OPENQASM 3;
include "stdgates.inc";

def bell(qubit[2] q) {
    h q[0];
    cx q[0], q[1];
}

def test_distribution() {
    // Configuration pragma (see Section 2.1)
    //% shots 10000
    //% seed 42
    //% backend ideal

    // Arrange
    qubit[2] q;
    bit[2] m;

    // Act
    bell(q);
    m = measure q;

    // Assert: Assertion Pragmas (see Section 2.2)
    //% expect distribution ref {"00": 0.5, "11": 0.5}
    //% expect distribution chi2 >= 0.05
}
\end{lstlisting}

\subsection{Configuration Pragmas}\label{sec:pragma:conf}

Configuration pragmas set the execution environment for a test function.
\textsc{QUTest} supports five such directives (e.g., lines 11--13 in
\Cref{lst:bell}).
\begin{enumerate}
  \item \textbf{Measurement shots:} \texttt{1024} (default).
    \begin{lstlisting}[style=qasm,numbers=none,aboveskip=1mm,belowskip=1mm,xleftmargin=0em,xrightmargin=0pt,escapeinside={(*@}{@*)},]
//% shots (*@\color{webgreen}$\langle N \rangle$@*)
    \end{lstlisting}

  \item \textbf{Simulator seed:} \texttt{random} (default).
    \begin{lstlisting}[style=qasm,numbers=none,aboveskip=1mm,belowskip=1mm,xleftmargin=0em,xrightmargin=0pt,escapeinside={(*@}{@*)}]
//% seed (*@\color{webgreen}$\langle N \rangle$@*)
    \end{lstlisting}

  \item \textbf{Backend:} \texttt{ideal} (default), \texttt{noisy}, or \texttt{hardware}.
  The \texttt{ideal} backend uses a noiseless statevector simulator.
  The \texttt{noisy} backend applies a depolarising noise model with a
  single-qubit error rate of $p_1 = 10^{-3}$ and a two-qubit error rate of
  $p_2 = 10^{-2}$, representative of current superconducting-qubit hardware.
  The \texttt{hardware} backend is reserved for future integration with
  cloud-based quantum processors.
    \begin{lstlisting}[style=qasm,numbers=none,aboveskip=1mm,belowskip=1mm,xleftmargin=0em,xrightmargin=0pt,escapeinside={(*@}{@*)}]
//% backend (*@\color{webgreen}$\langle B \rangle$@*)
    \end{lstlisting}

  \item \textbf{Runtime:} \texttt{qiskit}.
  The \texttt{runtime} pragma specifies which quantum software runtime
  executes the test.
  In the current implementation, \textsc{QUTest} supports only Qiskit.
    \begin{lstlisting}[style=qasm,numbers=none,aboveskip=1mm,belowskip=1mm,xleftmargin=0em,xrightmargin=0pt]
//% runtime qiskit
    \end{lstlisting}

  \item \textbf{Runtime version:} a quoted, comma-separated list of exact runtime versions.
  If this pragma is absent, the test runs once in the active Python
  environment.
  When it is present, the test is executed independently under each listed
  version.
    \begin{lstlisting}[style=qasm,numbers=none,aboveskip=1mm,belowskip=1mm,xleftmargin=0em,xrightmargin=0pt]
//% runtime_version "2.3.0,2.4.0"
    \end{lstlisting}
\end{enumerate}

\subsection{Assertion Pragmas}\label{sec:pragma:assert}

Assertion pragmas are central to \textsc{QUTest}. The framework currently
supports 12~assertion types organized into four categories (e.g., lines 24 and
25 in \Cref{lst:bell}). Although many additional assertion mechanisms have
been proposed, e.g., \cite{11421015}, implementing
all of them is beyond the scope of a prototype such as \textsc{QUTest}.

\subsubsection{Deterministic Output}

The \texttt{output} assertion checks whether every shot produces the same
classical value:
\begin{lstlisting}[style=qasm,numbers=none,aboveskip=1mm,belowskip=1mm,xleftmargin=0em,xrightmargin=0pt]
//% expect output m == 1
\end{lstlisting}
This assertion is the quantum analogue of a classical equality check and is
appropriate for circuits that deterministically prepare a computational-basis
state (e.g., a sequence of Pauli-X gates).

\subsubsection{Statistical Distribution}

For probabilistic circuits, \textsc{QUTest} supports assertions over the
measured output distribution, e.g., line~24 in \Cref{lst:bell}.
The following pragmas compare the empirical distribution against the provided reference
using distribution distances or statistical goodness-of-fit tests.
Each assertion accepts a comparison operator 
and a threshold.
\begin{enumerate}
  \item \textbf{Total variation distance:} $\delta(P, Q) = \frac{1}{2}\sum_x |P(x) - Q(x)|$.
  \begin{lstlisting}[style=qasm,numbers=none,aboveskip=1mm,belowskip=1mm,xleftmargin=0em,xrightmargin=0pt]
//% expect distribution total_variation_distance <= 0.03
  \end{lstlisting}

  \item \textbf{Hellinger distance:} $H(P, Q) = \frac{1}{\sqrt{2}}\sqrt{\sum_x (\sqrt{P(x)} - \sqrt{Q(x)})^2}$.
  \begin{lstlisting}[style=qasm,numbers=none,aboveskip=1mm,belowskip=1mm,xleftmargin=0em,xrightmargin=0pt]
//% expect distribution hellinger <= 0.05
  \end{lstlisting}

  \item \textbf{Kullback--Leibler divergence:} $D_{\mathrm{KL}}(P \| Q) = \sum_x P(x) \ln \frac{P(x)}{Q(x)}$.
  \begin{lstlisting}[style=qasm,numbers=none,aboveskip=1mm,belowskip=1mm,xleftmargin=0em,xrightmargin=0pt]
//% expect distribution kl <= 0.10
  \end{lstlisting}

  \item \textbf{Chi-squared goodness-of-fit test:}
  $\chi^2 = \sum_x \frac{(O_x - E_x)^2}{E_x}$, where $O_x$ is the observed
  count and $E_x = N Q(x)$ is the expected count under the reference
  distribution, and compares the resulting p-value.
  \begin{lstlisting}[style=qasm,numbers=none,aboveskip=1mm,belowskip=1mm,xleftmargin=0em,xrightmargin=0pt]
//% expect distribution chi2 >= 0.05
  \end{lstlisting}
\end{enumerate}

\subsubsection{Quantum-State Properties}

This category contains eight assertion types that probe different aspects of
the output.
\begin{enumerate}[leftmargin=*]
  \item \textbf{Marginal probability:} Assert the probability of a single qubit outcome.
    \begin{lstlisting}[style=qasm,numbers=none,aboveskip=1mm,belowskip=1mm,xleftmargin=0em,xrightmargin=0pt]
//% expect marginal m[0] p1 ~= 0.5 atol=0.05
    \end{lstlisting}

  \item \textbf{Pauli-$Z$ observable:} Compute $\langle Z_i Z_j \cdots \rangle$
    from measurement counts, with each bitstring contributing
    $(-1)^{\text{parity}}$ at the specified qubit positions.
    \begin{lstlisting}[style=qasm,numbers=none,aboveskip=1mm,belowskip=1mm,xleftmargin=0em,xrightmargin=0pt]
//% expect observable "Z0 Z1" ~= 1.0 atol=0.05
    \end{lstlisting}

  \item \textbf{Shannon entropy:} Assert the entropy (in bits) of the output distribution: $H = -\sum_x P(x) \log_2 P(x)$.
    \begin{lstlisting}[style=qasm,numbers=none,aboveskip=1mm,belowskip=1mm,xleftmargin=0em,xrightmargin=0pt]
//% expect entropy ~= 1.0 atol=0.05
    \end{lstlisting}

  \item \textbf{Pearson correlation:} Assert the classical correlation between
    two output bits, which ranges from
    $-1$ 
    to $+1$. 
    \begin{lstlisting}[style=qasm,numbers=none,aboveskip=1mm,belowskip=1mm,xleftmargin=0em,xrightmargin=0pt]
//% expect correlation m[0] m[1] ~= 1.0 atol=0.05
    \end{lstlisting}

  \item \textbf{Bitstring probability:} Assert the probability of a specific outcome without requiring a full reference distribution.
    \begin{lstlisting}[style=qasm,numbers=none,aboveskip=1mm,belowskip=1mm,xleftmargin=0em,xrightmargin=0pt]
//% expect probability "00" ~= 0.5 atol=0.05
    \end{lstlisting}

  \item \textbf{Most-frequent outcome:} Assert that the most frequently
    observed bitstring equals an expected value---a simple argmax check with
    no statistical test.
    \begin{lstlisting}[style=qasm,numbers=none,aboveskip=1mm,belowskip=1mm,xleftmargin=0em,xrightmargin=0pt]
//% expect most_frequent "00"
    \end{lstlisting}

  \item \textbf{Classical fidelity:} Compute the squared Bhattacharyya
    coefficient, $F = \bigl(\sum_x \sqrt{P(x) Q(x)}\bigr)^2$, between the
    measured distribution and an automatically obtained ideal simulation.
    \begin{lstlisting}[style=qasm,numbers=none,aboveskip=1mm,belowskip=1mm,xleftmargin=0em,xrightmargin=0pt]
//% expect fidelity >= 0.95
    \end{lstlisting}

  \item \textbf{Entanglement witness:} Verify that two qubit partitions are
    entangled by computing the von~Neumann entropy
    $S(\rho_A) = -\mathrm{tr}(\rho_A \log_2 \rho_A)$ of the reduced density
    matrix obtained via partial trace. Entanglement is confirmed when
    $S(\rho_A) > 0$.
    \begin{lstlisting}[style=qasm,numbers=none,aboveskip=1mm,belowskip=1mm,xleftmargin=0em,xrightmargin=0pt]
//% expect entangled [0] [1]
    \end{lstlisting}
\end{enumerate}

\subsubsection{Structural Assertions}

Are evaluated on the \emph{transpiled} circuit before execution, allowing developers to enforce hardware constraints without consuming simulator time.
\begin{itemize}[leftmargin=*]
  \item \textbf{Gate-set membership:} Assert that all gates in the transpiled circuit belong to a specified set.
    \begin{lstlisting}[style=qasm,numbers=none,aboveskip=1mm,belowskip=1mm,xleftmargin=0em,xrightmargin=0pt]
//% expect gateset subset_of [h, cx, rz]
    \end{lstlisting}

  \item \textbf{Depth:} Assert an upper bound on the transpiled circuit depth.
    \begin{lstlisting}[style=qasm,numbers=none,aboveskip=1mm,belowskip=1mm,xleftmargin=0em,xrightmargin=0pt]
//% expect depth <= 10
    \end{lstlisting}
\end{itemize}

\subsection{Comparison Operators}

Assertions support two comparison forms.
\emph{Approximate equality} (\verb|~=| \texttt{value atol=tolerance}) checks
$|actual - expected| \leq \texttt{atol}$ for marginals, observables,
entropy, correlation, probability, and fidelity.
Standard operators (\texttt{<}, \texttt{<=}, \texttt{==}, \texttt{>},
\texttt{>=}, \texttt{!=}) express exact or ordered checks for outputs,
distribution metrics, and depth bounds. Entropy supports both.


\section{QUTest}
\label{sec:architecture}

This section describes the \textsc{QUTest} framework.

\subsection{Discovery}

\textsc{QUTest} recursively scans a directory (or a single file) for
\texttt{.qasm} files and identifies all subroutines whose names begin with
\texttt{test}.

\subsection{Static Linting}\label{sec:architecture:lint}

Because \textsc{QUTest} encodes tests configuration and oracles as comments,
standard OpenQASM parsers cannot detect malformed pragmas.  To expose such
errors before execution, \textsc{QUTest} provides a static linter,
\texttt{qutest lint}.

The linter scans \texttt{.qasm} files without executing circuits. It ensures
that \texttt{//\%} directives appear inside \texttt{def~test*()} functions,
parses all directives, and reports line-level diagnostics with repair hints.
This mitigates a key drawback of comment-embedded pragmas: they preserve
OpenQASM compatibility but lack type checking.

\vspace{-5pt}
\subsection{Parsing and Function Inlining}\label{sec:architecture:parse}

Each \texttt{.qasm} file is parsed to extract function definitions.  Names
beginning with \texttt{test} are treated as tests; all others are considered
part of the software under test.  The pragma parser extracts \texttt{//\%} lines
from each test body and converts them into
objects.

At the momement, Qiskit's loader does not support OpenQASM~3 subroutines (\texttt{def}).
Thus, to address this limitation,
\textsc{QUTest} applies a \emph{function-inlining} pass before handing the source
to Qiskit.  Each call in a test body to an existing subroutine is replaced by the
callee body, substituting formal parameters with actual arguments.  This
process repeats transitively until all calls are expanded.

\vspace{-5pt}
\subsection{Environment-Aware Runtime Execution}\label{sec:architecture:env}

\textsc{QUTest} splits ordinary tests from environment-aware tests, and
groups the latter by
requested version. For each version, it creates or reuses a managed virtual
environment under \path{.qutest/runtimes/<runtime>/<version>/}, installs the
local \textsc{QUTest} package and the exact runtime version.

Before running any test, each managed environment must pass a
\emph{runtime compatibility probe}. The probe loads a minimal OpenQASM
circuit, selects a simulator backend, transpiles and executes the circuit,
and checks a deterministic oracle. If it fails, \textsc{QUTest} reports a
framework/runtime compatibility error rather than a user-test failure. In
effect, \texttt{runtime\_version} means ``run this test under that runtime
version if \textsc{QUTest} supports it.''

Each version group runs in a worker subprocess using that environment's
Python interpreter. If \texttt{runtime\_version} is present, ordinary tests also
also run in a subprocess under the active
interpreter, while the parent process handles discovery, environment setup,
worker invocation, and result aggregation. This isolation prevents import leakage
across environments. Any setup, probe, runtime-test, or worker failure makes
\textsc{QUTest} exit non-zero.

\vspace{-5pt}
\subsection{Execution}\label{sec:architecture:exec}

Each worker loads the inlined OpenQASM source
and
transpiles it for the configured backend with \texttt{transpile()}.
Structural assertions (gate-set membership and circuit depth) are checked
on the transpiled circuit before execution.
It then runs the circuit on the selected \texttt{AerSimulator} with the
configured shots and optional seed.
Statevector-dependent tests (e.g., the entanglement witness) use a
separate simulation of a measurement-free copy.

\vspace{-5pt}
\subsection{Assertion Evaluation}\label{sec:architecture:assert}

Each assertion pragma has a dedicated evaluator that receives
measurement counts, a transpiled circuit, a statevector, or ideal counts,
and returns an \texttt{AssertionResult} with a status (\texttt{pass},
\texttt{fail}, or \texttt{error}), a message, and the
actual and expected values.

\vspace{-5pt}
\subsection{Reporting}\label{sec:architecture:report}

\textsc{QUTest} supports two reporting modes.
  \textbf{Console reporter:} prints coloured checkmarks
    (\texttt{\checkmark}) for passing tests and crosses
    (\texttt{\texttimes}) for failures. In verbose mode, all assertion
    details are shown; in the default mode, only failing assertions are
    expanded.
  \textbf{XML reporter:} generates an XML file following the
    JUnit/xUnit schema, suitable for CI systems such as Jenkins, GitHub~Actions,
    and GitLab~CI.

\vspace{-5pt}
\subsection{CLI}

\textsc{QUTest} is a Python~3 package built on Qiskit~$\geq$1.0,
Qiskit~Aer~$\geq$0.13, \texttt{openqasm3}, and
\texttt{qiskit-qasm3-import}, with NumPy and SciPy for numerics.
Its CLI exposes \texttt{lint} for the static
checks in \Cref{sec:architecture:lint} and \texttt{run} for the execution
workflow in
\Cref{sec:architecture:parse,sec:architecture:env,sec:architecture:exec,sec:architecture:assert,sec:architecture:report}.

\vspace{-5pt}
\subsection{Design Trade-offs}

\noindent
\textbf{Pragmas vs.\ language extension.}
Rather than adding native OpenQASM~3 assertions, \textsc{QUTest} uses comment
pragmas: no grammar or parser changes and compatibility with existing tools,
at the cost of weaker tooling (no type checking, limited IDE support). The
linter and a language-server extension could help.

\noindent
\textbf{Inlining vs.\ native support.}
Because the current Qiskit importer cannot load \texttt{def} subroutines,
\textsc{QUTest} inlines them before execution on the runtime. Once parsers
support \texttt{def}, this pass can disappear. The current inliner supports
simple parameter substitution, but not nested-scope name collisions.

\vspace{-5pt}
\section{Planned Evaluation}\label{sec:evaluation}

We plan to evaluate \textsc{QUTest} along two dimensions: developer effort
and test effectiveness.

\noindent
\textbf{Developer Effort.}
The first study would assess whether developers can write QASM-native tests
with \textsc{QUTest} more easily than Python-based harnesses.
In a within-subject design, participants would test small OpenQASM programs
from natural-language tasks, with task order counterbalanced.
We would collect objective measures (e.g., completion time, number of files and lines of test code, edit-run-debug cycles, and syntax/oracle mistakes) and subjective
measures from questionnaires (e.g., perceived difficulty, confidence,
readability, and required host-SDK familiarity). Together, these data would
show whether QASM-native pragmas reduce test-writing effort or introduce
usability costs of their own.

\noindent
\textbf{Test Effectiveness.}
The second study would evaluate the adequacy of
Python-based
and \textsc{QUTest} test suites at exercising quantum programs structure~\cite{fortunato2026probabilisticCoverage} and detecting faults~\cite{QMutPy,Muskit}.  This will allow us to assess whether reduced test-writing effort weakens fault detection or
whether QASM-native tests enable more effective oracles.

\vspace{-5pt}
\section{Discussion}
\label{sec:discussion}

\paragraph{Python-Based Testing vs.\ \textsc{QUTest}}
Comparing \Cref{lst:intro-bell-qasm,lst:intro-bell-pytest} with
\Cref{lst:bell} highlights the difference in development overhead. The former requires (a) two languages (OpenQASM and Python) and three
Python libraries (Qiskit, SciPy, and PyTest), whereas the latter requires only
OpenQASM and a single CLI tool (\textsc{QUTest});
and (b) uses $7 + 18 = 25$ lines to implement the program under
test and one single-assertion test case. The latter uses 21 lines for the same
program and test.
Beyond the listing, \textsc{QUTest} also allows a single test to target multiple runtime environments (see \Cref{sec:pragma:conf}) and express multiple assertions as one-line pragmas (see \Cref{sec:pragma:assert}).

\emph{Who is \textsc{QUTest} for?}
\textbf{Researchers and developers} who write
OpenQASM and do not want---or may not know how---to write test code in a
quantum SDK just to verify expected behavior.
\textbf{SDK maintainers.} With the \texttt{runtime} and
\texttt{runtime\_version} pragmas (see \Cref{sec:architecture:env}),
\textsc{QUTest} treats the execution environment as part of the test
specification. The same QASM test can thus be run across Qiskit versions to
expose behavioral drift. Examples include the QAOA migration regression
reported by \citet{cardinal2025qaoaMigration} and Qiskit issue~\#12124~\cite{qiskitIssue12124},
where \texttt{qiskit.qasm2.loads} failed to recognize gates from
\texttt{qelib1.inc}.

\vspace{-5pt}
\section{Related Work}
\label{sec:related}

QuCAT~\cite{qucat} and QuSBT~\cite{qusbt} generate Python-based test
inputs, and QuraTest~\cite{quratest} and NovaQ~\cite{novaq} generate OpenQASM
test circuits.
These tools are complementary to \textsc{QUTest}: they generate inputs,
whereas \textsc{QUTest} provides a QASM-native format for complete test cases,
including execution configuration and assertions. In particular, generated
inputs still need a harness to load, execute, and check expected behavior;
\textsc{QUTest} provides that missing layer. Future tools could combine both by
generating native \textsc{QUTest} tests.

\vspace{-5pt}
\section{Conclusion and Future Work}
\label{sec:conclusion}

We have presented \textsc{QUTest}, the first framework that enables quantum developers to write test directly in OpenQASM~3.
Future work includes 
\textbf{parameterised tests},
\textbf{support for other quantum SDKs}, e.g., Cirq~\cite{cirq} and Pennylane~\cite{pennylane}, and \textbf{real hardware backends}.

\bibliographystyle{ACM-Reference-Format}
\bibliography{main}

\end{document}